\documentclass[]{raa}            
\usepackage{graphicx,times}
\usepackage{natbib}

\usepackage{amsmath}
\usepackage{amssymb}
\usepackage{multirow}
\usepackage{xcolor}

\usepackage{setspace}
\providecommand{\bm}[1]{\mbox{\boldmath$#1$\unboldmath}}

\begin{document}

   \title{Merging strangeon stars II: the ejecta and light curves
}

\volnopage{ {\bf 2021} Vol.\ {\bf X} No. {\bf XX}, 000--000}
   \setcounter{page}{1}

   \author{Xiao-Yu Lai\inst{1,2}, Cheng-Jun Xia\inst{3}, Yun-Wei Yu\inst{4}, Ren-Xin Xu\inst{5,6}
   }

   \institute{Department of Physics and Astronomy, Hubei University of Education, Wuhan 430205, China; {\it laixy@hue.edu.cn}\\
        \and
             Research Center for Astronomy, Hubei University of Education, Wuhan 430205, China\\
    \and
       School of Information Science and Engineering, NingboTech University, Ningbo 315100, China\\
   \and
     Institute of Astrophysics, Central China Normal University, Wuhan 430079, China\\
     \and
     School of Physics, Peking University, Beijing 100871, China\\
     \and
     Kavli Institute for Astronomy and Astrophysics, Peking University, Beijing 100871, China\\
      \vs \no
   {\small Received [year] [month] [day]; accepted [year] [month] [day]}
}

\abstract{The state of supranuclear matter in compact stars remains
puzzling, and it is argued that pulsars could be strangeon stars.
The consequences of merging double strangeon stars are worth
exploring, especially in the new era of multi-messenger astronomy.
To develop the ``strangeon kilonova'' scenario proposed in Paper I,
we make a qualitative description about the evolution of ejecta and
light curves for merging double strangeon stars.
In the hot environment of the merger, the strangeon nuggets ejected
by tidal disruption and hydrodynamical squeezing would suffer from
evaporation, in which process particles, such as strangeons,
neutrons and protons, are emitted.
Taking into account both the evaporation of strangeon nuggets and
the decay of strangeons, most of the strangeon nuggets would turn
into neutrons and protons, within dozens of milliseconds after being
ejected.
The evaporation rates of different particles depend on temperature,
and we find that the ejecta could end up with two components, with
high and low opacity respectively.
The high opacity component would be in the directions around the
equatorial plane, and the low opacity component would be in a broad
range of angular directions.
The bolometric light curves show that even if the total ejected mass
would be as low as $\sim 10^{-4} M_\odot$, the spin-down power of
the long-lived remnant would account for the whole emission of
kilonova AT2017gfo associated with GW 170817.
The detailed picture of merging double strangeon stars is expected
to be tested by future numerical simulations.
 }

   \authorrunning{X.-Y. Lai et al. }            
   \titlerunning{Merging strangeon stars II}  
   \maketitle

%

\section{Introduction}
\label{Introduction}

Matter in our Universe takes on various forms, although the
fundamental particles making up matter are just three generations of
Fermions in the Standard Model of particle physics.
The state of matter at extremely high densities created by the
gravitational collapse of massive stars is still far from certainty,
which is yet essential for us to explore the nature of pulsar-like
compact stars.
It is still under debate if the main constitution of pulsar-like
compact stars is two-flavored or three-flavored matter.
The gravity-compressed matter produced after a core-collapse
supernova of an evolved massive star is currently speculated be
either neutron matter or strange matter, and a historical roadmap to
these ideas is introduced briefly by~\citet{Xu2021}.

For bulk matter, at densities around the saturated nuclear matter
density $\rho_0$, the weak equilibrium among u, d and s quarks is
possible, instead of simply that between u and d quarks.
Rational thinking about stable strangeness dates back to 1970s.
Bulk strange object, composed of nearly equal numbers of u, d and s
quarks, is speculated to be the absolutely stable ground state of
strong-interacting matter, which is known as Witten's
conjecture\citep{Witten1984}.
It should be also noted that, due to the non-perturbative effect of
strong interaction, quarks inside pulsar-like compact stars may be
grouped into clusters, similar to the case that u and d quarks are
grouped into nucleons.
Although Witten's conjecture was proposed based on {\it strange
quark matter} that composed of almost free quarks, we can make an
extension that it still reasonably holds no matter whether quarks
are free or localized.

At densities in compact stars, which are around the saturated
nuclear matter density, the coupling between quarks would be so
strong that quarks are hard to maintain itinerant.
Initiated by the thoughts that the quark-clusters being the main
constituents of compact stars in~\citet{Xu2003} and~\citet{LX2009a},
where the three-flavored quark-clusters are afterward called
``strangeons'' by combining ``strange nucleons'', this model has
been developed based on more advanced observations (see a review
by~\citet{LX2017} and references therein).

The strangeon star model had been found to be helpful to understand
different manifestations of pulsar-like compact stars.
Strangeon star model predicts high mass
pulsars~\citep{LX2009a,LX2009b} before the discovery of pulsars with
$M>2M_\odot$~\citep{Demorest2010}.
The strangeon matter surface could naturally explain the pulsar
magnetospheric activity~\citep{Xu1999ApJL} as well as the
subpulse-drifting of radio pulsars~\citep{Lu2019}.
Starquakes of solid strangeon stars could induce
glitches~\citep{Zhou2004,Zhou2014,Lai2018MN}, and the relation
between the recovery coefficients and glitch sizes was found to be
consistent with observations~\citep{Lai2018MN}.
The glitch activity of normal radio
pulsars~\citep{Lyne2000,Espinoza2011,Fuentes2017} can also be
explained under the framework of starquake of solid strangeon star
model~\citep{WangWH2020}.
The plasma atmosphere of strangeon stars can reproduce the
Optical/UV excess observed in X-ray dim isolated neutron
stars~\citep{Kaplan2011,Wang2017APJ}.
The tidal deformability~\citep{LZX2019} as well as the light
curve~\citep[hereafter Paper I]{Lai2018RAA} of merging binary
strangeon stars have been derived, which are consistent with the
results of gravitational wave event GW170817~\citep{GW170817}, and
the details will be explained later.

The inner structure of pulsar-like compact stars as well as the
equation of state (EOS) of supranuclear dense matter are challenging
in both physics and astronomy.
The significant non-perturbative effect makes it difficult to derive
the properties of dense matter inside pulsar-like compact stars from
the first principle.
The theoretical models (including neutron star model as the
mainstream, quark star model and strangeon star model) need to be
tested by the astrophysical observations.

{\it Strangeon matter}, similar to {\it strange quark matter}, are
composed of nearly equal numbers of u, d and s quarks at the level
of quarks; however, different from that in {\it strange quark
matter}, quarks in {\it strangeon matter} are localized inside
strangeons due to the strong coupling between quarks.
There are differences and similarities between strangeon stars and
neutron/quark stars.
On the one hand, quarks are thought to be localized in strangeons in
strangeon stars, like neutrons in neutron stars, but a strangeon has
3 flavors and may contain more than three valence quarks.
On the other hand, the matter at the surface of strangeon stars is
still strangeon matter, i.e., strangeon stars are self-bond by
strong force, like quark stars.

The detections of gravitational wave event GW170817~\citep{GW170817}
and its multiwavelength electromagnetic
counterparts~\citep[e.g.,][]{Kasliwal2017,Kasen2017} open a new era
in which the nature of pulsar-like compact stars could be crucially
tested.
In the conventional neutron star merger, the neutron-rich ejecta
undergoes rapid neutron capture (r-process) nucleosynthesis.
The radioactive decay of these unstable nuclei powers a rapidly
evolving and supernova-like transient named as AT2017gfo, which was
predicted to be associated with neutron star mergers and in
literatures was called ``kilonova''~\citep{Li1998},
``macronova''~\citep{Kulkarni2005}, or
``mergernova''~\citep{Yu2013,Gao2015}.
The observed multi-band light curves can be understood by such
radioactivity-powered
transient~\citep[e.g.,][]{Cowperthwaite2017,Smartt2017,Villar2017},
containing a low-opacity ($\kappa \sim 10^{-1}$ cm$^2$ g$^{-1}$)
component (``blue'' component) whose luminosity peaks at $\sim
10^{42}$ erg s$^{-1}$ at the time about one day, and a high-opacity
($\kappa \sim 10-100$ cm$^2$ g$^{-1}$) component (``red'' component)
whose luminosity peaks at the time about one week.

Combining the constraint by GW170817 with the existence of high mass
pulsars puts a dramatic reduction in the family of allowed equation
of states of neutron stars~\citep{Annala2017}.
As more massive pulsars are being
found~\citep{Demorest2010,Antoniadis2013,Cromartie2019}, the lower
limit of the maximum mass increases, which will put more stringent
constraint on neutron star model.
Due to the lack of information on the post-merger remnant, the
observation gravitational wave alone cannot exclude other
possibilities of the origin of GW170817.
For binary quark stars, the tidal deformability of GW170817 can be
used to constrain parameters in the equation of state, which imply
the maximum mass of quark stars to be $\sim 2.18 M_\odot$ within the
MIT bag model~\citep{Li2017} and $\sim 2.32 M_\odot$ with
color-flavor-locked superfluity~\citep{Li2020}.

Differently, for equation of state of strangeon stars, the
constraint by combining the tidal deformability and high maximum
mass seems not severe at all\footnote{The strangeon star model is
neither in the ``twin-stars'' scenario~\citep{Most2018} nor in the
``two-families'' scenario~\citep{Drago2019a}.}.
For the strangeon matter proposed in~\citet{LX2009b}, in a large
parameter space the equation of state of strangeon star is
compatible with the constraint by GW170817 even if the maximum mass
of pulsars is higher than 2.8 $M_\odot$~\citep{LZX2019}.
For the linked bag model of strangeon matter~\citep{Miao2020} which
can be adopted for strong condensed matter in both 2-flavoured
(nucleons) and 3-flavoured (hyperons and strangeons) scenarios, it
is also found that in a large parameter space the maximum mass and
tidal deformability of strangeon stars are consistent with the
current astrophysical constraints.

It is interesting to note that, some studies showed that, when
introducing realistic current quark masses, the strange quark
becomes disfavored because of its large dynamical mass, and the
three flavored strange quark matter would not be absolutely
stable~\citep{Buballa1998}.
Under such consideration, quark matter with only u and d quarks
(udQM) has also gained some attentions.
By some phenomenological models for interacting quarks, udQM was
shown to be more stable than nuclear matter and strange quark
matter~\citep{Holdom2017}.
The maximum mass of quark stars with udQM could be larger than
$2.7M_\odot$~\citep{Cao2020}, and the obtained values of the tidal
deformability are in good compatibility with the experimental
constraints of GW170817~\citep{Zhang2020}.
Therefore, it remains an interesting and unsolved problem that
whether the quark matter is 2- or 3-flavored.
Strangeon matter that we focus on in this paper is also the result
of significant interaction between quarks.

Besides determining the tidal deformability, the equation of state
of compact stars also determines the properties of post-merger
remnant, which would affect the electromagnetic transient after
merger.
The allowed equations of state of neutron stars and strange quark
stars are hard to sustain a mass higher than $2.5
M_\odot$~\citep{Annala2017,Li2017}, so the remnant of merger for
GW170817 is more likely to be short-lived and will be collapse into
a black hole within 100 ms~\citep{Ruiz2017}.
The lanthanide-bearing ejecta is important for the ``red'' component
of the post merger light curves, but most of the ejecta is
lanthanide-free ($Y_e\gtrsim 0.3$) if the neutron star survives
longer than about 300 ms~\citep{Kasen2015}.
However, a long-lived neutron star is favored for a consistent
picture to account for the opacity and ejected mass of
AT2017gfo~\citep{Yu2017,LiYu2018}.

The observed electromagnetic counterparts, on the other hand, are
still difficult to directly probe the nature of pulsar-like compact
stars.
The production of heavy elements has impact on the opacity and will
consequently affect the time and magnitude of peak luminosity.
Neutron star mergers could not be the only complement to supernovae
that produce elements around or heavier than iron peak.
Merger of double quark stars would eject fragments of strange quark
matter, which are called strangelets.
For mergers of double quark stars, under the multi-fragmentation
model~\citep{Horvath2014} of quark matter, all the ejected
strangelets would decay into nuclear matter, and the nucleosynthesis
of quark star mergers would reach the iron peak
only~\citep{Paulucci2017}.
\citet{Drago2019b} calculate the evaporation process of ejected
strangelets, and find that almost all of the ejected strangelets
will evaporate into nucleons (most of them are neutrons).
Although the evaporation of strangelets into nucleons could produce
neutron-rich condition and then could lead to high opacity, there is
a lack of explanation about the observed low opacity component.

The consequences of merging double strangeon stars are worth
exploring.
The ``strangeon kilonova'' scenario has been discussed in Paper I,
in which the peak of the light curve at about one day after merger
is powered by the decay of ejected unstable strangeon nuggets, and
the slowly fading component of the light curve is powered by the
spin-down of the remnant strangeon star\footnote{Such a hybrid
energy source model was firstly suggested by~\citet{Yu2017} for
explaining AT 2017gfo with a long-lived normal neutron star.}.
To match the observations, the lifetime of the unstable strangeon
nuggets is assumed to be one day.
However, the detailed descriptions about the evolution of ejected
strangeon nuggets as well as the properties of the decay products is
needed.

To present a whole picture of merging double strangeon stars and the
astrophysical consequences, there is still a long way to go.
The full analysis about ejection process of strangeon nuggets,
including the total mass and the size-distribution of nuggets,
relies on numerical simulations.
In addition, the evolution of ejected strangeon nuggets is difficult
to trace due to our ignorance of their properties.
However, as will be shown in this paper, the ejection and evolution
of strangeon nuggets happened and terminated at very early stage of
merger, so these processes could not have much impact on the later
processes such as the strangeon kilonova.
As a first stage exploring the astrophysical consequences of merging
double strangeon stars, a qualitative description about the
evolution of ejecta and light curve of kilonova is necessary, which
is focused in this paper.

Beginning with a rough picture for ejection of strangeon nuggets
during merger of double strangeon stars in
Section~\ref{sec:ejection}, we discuss the evaporation of ejected
strangeon nuggets in Section~\ref{sec:evaporation}.
It is found that, except the ones that have initial baryon numbers
near the maximum value, almost all the ejected nuggets turn into
strangeons at within several milliseconds, and turn into neutrons
and proton within tens of milliseconds.
Because strangeons would instantly decay that leads to more protons
than neutrons, the high and low opacity components could be
naturally created.
Although the total ejected mass of would be as low as $\sim
10^{-4}$, the light curve would be powered by the spin-down of the
remaining long-lived strangeon star, which can fit the bolometric
light curve of AT2017gfo, as shown in Section~\ref{sec:strangeon
kilonova}.
Conclusions and discussions are made in
Section~\ref{sec:conclusions}.

\section{Ejection of strangeon nuggets}
\label{sec:ejection}

The electromagnetic counterparts of GWs in merging binary compact
stars are essentially determined by the amount and composition of
the ejecta.
Similar to quark stars, strangeon stars are self-bound on the
surface.
It is known that modelling the large discontinuities at the surface
of quark stars faces numerical challenges, and only a few works have
explored the dynamics of binary quark stars.
The hydrodynamical simulations of the coalescence of quark
stars~\citep{Bauswein2010} show that the small lumps of quark matter
form around the remnant, and the total ejected mass is $\sim 0.004
M_{\odot}$.
Recently, the fully general-relativistic simulations of binary quark
stars have been presented~\citep{ZhuZY2021}, which show that the
dynamical mass loss is significantly suppressed to be about $\sim
10^{-4} M_\odot$.

The clumpy ejecta and the low ejected mass are due to the fact that
quark stars are self-bound by the strong interaction, which is also
the character of strangeon stars.
We may expect that the ejected mass of merging binary strangeon
stars is more or less the same as that of merging quark stars,
although the full numerical simulations of binary strangeon stars
remains to be done.
In the following, we assume that there are two main ejection process
in the merger of double strangeon stars, similar to the merger of
quark stars.
The first process is the tidal disruptions during the merger,
ejecting matter in the equatorial plane.
The second process is the hydrodynamical squeeze from the contact
interface between the merging stars, expelling matter in a broad
range of angular directions.
We further assume that the ejected matter in both processes has mass
as low as $\sim 10^{-4} M_{\odot}$.

Due to the self-binding of strangeon stars, both tidal disruption
and hydrodynamical processes eject strangeon nuggets, instead of
ejecting individual strangeons.
The ejected strangeon nuggets could be like the water drops splashed
out of a pool of water, and they should have various sizes, i.e.
various baryon numbers $A$.
Here we can estimate the maximum and minimum sizes of strangeon
nuggets.

The maximum size could be estimated by the balance between tidal
force $GMmr/R^3$ and surface tension force $\sigma r$, where
$\sigma$ is the surface tension, $M$ and $R$ are the stellar mass
and radius, $m$ and $r$ are the nugget's mass and radius ($m\sim\rho
r^3$, $\rho$ is the density for both strangeon stars and strangeon
nuggets, $\rho\sim 2\rho_0$).
For $\sigma=10$ MeV fm$^{-2}$, we can get the maximum radius of
nuggets $r_{\rm max}\sim 1$ cm, corresponding to maximum baryon
number $A_{\rm max}\sim 10^{39}$.

From the method used in~\citet{Drago2019b}, the minimum size could
be estimated by evaluating the Weber number, defined by $W=\rho v^2
r/\sigma$, where $v$ is the turbulent velocity.
The ejection of strangeon nuggets can be treated as the turbulent
fragmentation on the surface of merging stars, where the turbulent
velocity $v$ is the ejection velocity.
The ejection takes place as long as $W\geq 1$, then if $v=0.1c$ ($c$
is the speed of light) we can get the minimum radius of nuggets
$r_{\rm min}\sim 1$ fm, corresponding to minimum baryon number
$A_{\rm min}\sim 1$.

Actually, the strangeon nuggets stable at zero temperature should
have a critical size, smaller than which the energy per baryon of
strangeon matter would be higher than that of two-flavor ordinary
matter\footnote{Note the difference between strangeon nuggets and
strangeons. The former are composed of the latter. A strangeon
nuggets would be stable against decaying to two flavor matter if its
baryon number is higher than the critical value (a strangeon star is
a huge ``nugget'' with baryon number $\sim 10^{57}$). A strangeon
has baryon number $\lesssim 10$ and is extremely unstable in vacuum
(their decay would lead to interesting consequences, discussed
in~\ref{subsec:electron fraction}).}.
In a qualitative estimation~\citep{LX2017} the critical size could
be set to be the Compton wavelength of electrons, $\lambda_{\rm
e}\sim 10^3$ fm, corresponding to the critical baryon number $A_{\rm
c}\sim 10^9$.
Then the primary strangeon nuggets that ejected during merger would
have baryon numbers from $10^9$ to $10^{39}$.

The lack of numerical simulations about the merging processes of
binary strangeon stars makes it hard to derive exactly the
distribution of sizes and the total amount of ejecta.
In fact, the size-distribution of strangeon nuggets would not have
much impact on the electromagnetic radiation.
After being ejected, strangeon nuggets will suffer evaporation, as
will shown in Sec~\ref{sec:evaporation}, and the the final
components in the ejecta which have observational effects (e.g. the
power of the kilonova) would depend weakly on the initial
conditions.

\section{Evaporation of strangeon nuggets}
\label{sec:evaporation}

During merger, the temperature could reach up to tens of
MeV~\citep[e.g.][]{Bauswein2010}, especially when the shock heating
is taken into account~\citep{Drago2019a}, so naturally the strangeon
nuggets would suffer from losing particles from the surface.
Strangeon nuggets themselves would behave like dark matter because
of their extremely low charge to mass ratio~\citep{LX2010}, but the
particles emitted from their surface would lead to significant
consequences.
In the high temperature environment of the merger, the ejected
strangeon nuggets would suffer from emission of particles from the
surface, i.e. evaporation, including neutrons, protons, strangeons,
and so on.

In this section, we calculate the evaporation rate of strangeon
nuggets, which depends on temperature.
It will be found that, the components in the ejecta after tens of
milliseconds from the merger would be similar to that in merging
double neutron stars.

\subsection{Widths of particle emissions}
\label{subsec:decay widths}

The width $\Gamma_\beta$ for the emission of particles can be
obtained with a statistical model~\citep{Sheng2005}, i.e.,
\begin{equation}
\Gamma_\beta(\varepsilon^*) = \frac{g_\beta m_\beta}{\pi^2}
\int_0^{\varepsilon^*-s_\beta}
\frac{\rho(\varepsilon^*-s_\beta-\varepsilon)}{\rho(\varepsilon^*)}
\varepsilon \sigma_\beta(\varepsilon) \mbox{d}\varepsilon,
 \label{eq:width}
\end{equation}
where $g_\beta$, $m_\beta$, and $s_\beta$ are the degeneracy factor,
mass, and separation energy of the particle.
Here $\rho(E^*)$ represents the level density of the a strangeon
nugget with an excitation energy $\varepsilon^*$, and
$\sigma_\beta(\varepsilon)$ the absorption cross section of particle
$\beta$ with an incident energy $\varepsilon$.
For electric neutral particles such as strangeons or neutrons we
take the cross section as $\sigma_{q,n}(\varepsilon) = \pi r^2$,
while for charged particles one has to take into account the Coulomb
interaction~\citep{Wong1973_PRL31-766}, i.e.,
\begin{equation}
\sigma_\beta(\varepsilon) = \frac{r^2 \omega_0}{2\varepsilon}
\ln{\left\{ 1 +
\exp{\left[\frac{2\pi(\varepsilon-\varepsilon_\mathrm{C})}{\omega_0}\right]}\right\}},
\label{eq:sigma_c}
\end{equation}
where the transmission probability of Coulomb barrier is obtained
based the Hill-Wheeler formula~\citep{Hill1953_PR089-1102} assuming
a typical barrier width $\omega_0 = 4$ MeV.
For the Coulomb barrier, we simply take $\varepsilon_\mathrm{C} =
q_\beta\varphi(r)$.

Note that for the emission of nucleons and $\alpha$ particles, one
needs to take into account the transition probability from
strangeons into nucleons.
If strangeon matter is usually more stable than nuclear matter and
the transition is a weak reaction process, we expect a vanishing
transition probability.
If we assume the transition probability from strangeons into
nucleons is $f_N$, the transition probability from strangeons into
$\alpha$ particles is approximately $f_N^4$.
Thus the cross sections become
\begin{equation}
  \sigma_{p,n}\rightarrow f_N\sigma_{p,n}\text{ and }\sigma_{\alpha}\rightarrow f_N^4\sigma_{\alpha}.
\end{equation}
At this moment, it is unclear the exact form of $f_N$.
According to the reaction rate of $s + u \rightarrow u + d$ in quark
matter given in~\citet{Madsen1993}, we suppose $f_N = 3\times
10^{-12}$.

When the temperature of strangeon matter exceeds certain value
($\sim 1$ MeV), the solid sate is turned into liquid.
In such cases, we expect the statistical properties of strangeon
nuggets are similar to finite nuclei, thus for strangeon nuggets we
adopt the level density of nuclei typically calculated from the
Fermi-gas model,
\begin{equation}
  \rho = \frac{{\rm e}^{2 \sqrt{a \varepsilon^*}}}{\sqrt{48}\varepsilon^*}, \label{eq: rho_NM}
\end{equation}
where the temperature is given by $T = \sqrt{\varepsilon^*/a}$.
The level density parameter is taken as $a = A/12A_q\
\mathrm{MeV}^{-1}$ since the effective degree of freedom is
strangeon instead of nucleon, where $A$ is the total baryon number
of a strangeon nugget, and $A_q$ is the baryon number of each
strangeon.
If the number of valence quarks in each strangeon is $N_q$, then the
strangeon baryon number $A_q=N_q/3$.
In this work we take $N_q=18$, i.e. $A_q=6$, in which case a
strangeon is an 18-quark cluster (called
quark-$\alpha$)~\citep{Michel1988}.

At large excitation energies ($\varepsilon^*\gg
s_\beta+\varepsilon$), the ratio of level densities in
Eq.~(\ref{eq:width}) can be simplified and gives
\begin{equation}
\frac{\rho(\varepsilon^*-s_\beta-\varepsilon)}{\rho(\varepsilon^*)}
=  \exp\left(-\frac{s_\beta+\varepsilon}{T}\right).
 \label{eq:width0}
\end{equation}

\subsection{Evaporation rate of strangeon nuggets}
\label{subsec:evporation rate}

Here we consider four evaporation channels, i.e., the emission of
strangeons ($\beta = q$), neutrons ($\beta = n$), protons ($\beta =
p$), and $\alpha$ particles ($\beta = \alpha$).
The separation energy is then obtained with $s_q = m_q - A_q \mu_n$,
$s_n = m_n - \mu_n$, $s_p = m_p - \mu_p$, and $s_\alpha = m_\alpha -
2\mu_n - 2\mu_p$, where $m_q$, $m_n$, $m_p$ and $m_\alpha$ are the
masses of strangeons, neutrons, protons and $\alpha$ particles,
respectively.
Here the neutron and proton chemical potentials are obtained with
$\mu_n = \frac{\partial M}{\partial A}$ and $\mu_p = \frac{\partial
M}{\partial Z}$, where $M$ is the mass of a strangeon nugget with
baryon number $A$ and charge number $Z$.

The emission rates $W_\beta = \Gamma_\beta/\hbar \approx 1.52\times
10^{21} \Gamma_\beta$ (in s$^{-1}$) for various evaporation channels
can be derived, where the widths $\Gamma_\beta$ are obtained with
Eq.~(\ref{eq:width}).
In principle, the surface tension $\sigma$ determining the dynamic
stability of the strangeon-vacuum interface would affect the
emission rate.
However, for larger strangeon nuggets (radius $r>10^5$ fm, or baryon
number $A>10^{15}$), the finite size effect becomes insignificant
and the emission rate is proportional to the surface area of
strangeon nuggets.
As will be shown in Sec~\ref{subsec:fate}, the strangeon nuggets
with initial baryon number $A_0\leq 10^{36}$ will almost disappear
within $\sim 1$ ms as the result of evaporation.
For simplicity we only consider large strangeon nuggets and neglect
the surface tension, since larger strangeon nuggets would emit more
particles.

In Fig.1 we present the emission rates per surface area for various
evaporation channels $\mathcal{R}_{\beta}$, including strangeons
($\beta=q$), neutrons ($\beta=n$), protons ($\beta=p$) and $\alpha$
($\beta=\alpha$).
The evaporation channels are dominated by strangeons at temperature
$T>10$ MeV, and dominated by neutrons at $T<10$ MeV.
This result gives the emission rates for any strangeon nuggets with
radius $r\gtrsim10^5$ fm via multiplying it by the surface area
$S=4\pi r^2$.
%
   \begin{figure}[h!!!]
  \centering
   \includegraphics[width=9.0cm, angle=0]{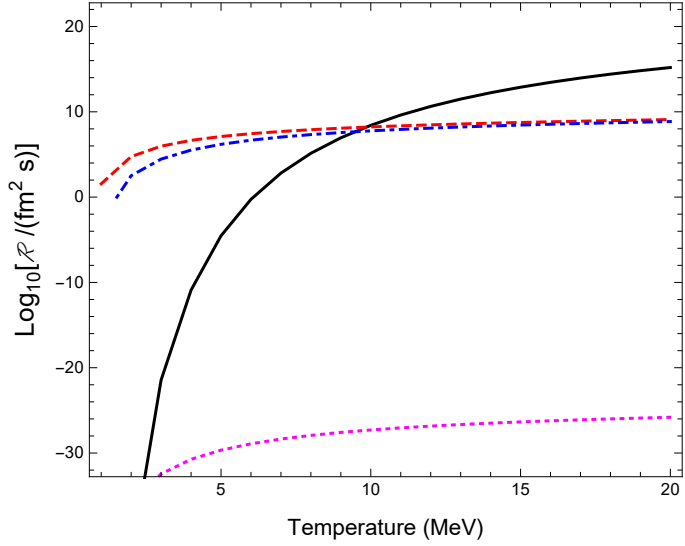}
   \begin{minipage}[]{85mm}
   \caption{Emission rates per surface area for evaporation channels $\mathcal{R}_{\beta}$ to strangeons (solid black line),
   neutrons (long-dashed red line), protons (dash-dotted blue line) and $\alpha$ particles (short-dashed magenta line) respectively.}
   \end{minipage}
   \label{fig_W}
   \end{figure}

The dependence of evaporation channels on temperature could be
understood.
Strangeons are heavier than neutrons and protons, so they are easier
to be emitted at high temperature.
If temperature is not high enough, it is energetically favored for
strangeons to decay into neutrons and protons before being emitted.
The emission of protons is suppressed due to the Coulomb barrier.

\subsection{The fate of strangeon nuggets}
\label{subsec:fate}

By simplifying the expanding envelope surrounding the remnant to be
of adiabatic~\citep{Li1998}, the temperature deceases as time,
$T\propto t^{-1}$.
As indicated in Fig.1, the production rate of strangeons, neutrons
and protons depends on the temperature.
If the initial temperature is $\sim 10$ MeV at the initial time
$\sim 1$ ms, then in $\sim 10$ ms the temperature decreases to 1
MeV, when the evaporation nearly ceases.
Therefore, evaporation only happens at very early stage of
expansion.

The calculations in Sec~\ref{subsec:evporation rate} show that at
different temperatures, the dominate evaporation products are
different.
When the temperature $T\simeq$ 20 MeV, the main evaporation products
are strangeons, with the rate (per unit surface area)
$\mathcal{R}=\mathcal{R}_q\simeq 1.5\times 10^{15}$ s$^{-1}$
fm$^{-2}$.
When the temperature is between 5 to 10 MeV, the main evaporation
products are neutrons, with the rate (per unit surface area)
$\mathcal{R}=\mathcal{R}_n\sim 10^8$ s$^{-1}$ fm$^{-2}$.
Then we can estimate the upper limit of initial baryon number $A$ of
strange nuggets which would almost disappear as the result of
evaporation.

Assuming each strangeon nugget is a sphere with radius $r$ and
baryon number density $n$.
When suffering evaporation, the rate of losing baryons from the
surface is
\begin{equation}
\frac{d A}{d t}=-\mathcal{R} \cdot 4\pi r^2,
\end{equation}
where
\begin{equation}
A=\frac{4\pi}{3}r^3n
\end{equation}
\begin{equation}
A\simeq \left(A_0^{1/3}-\frac{\mathcal{R}}{\rm fm^{-2}}\cdot t
\right)^3.\label{A}
\end{equation}
When $t=1$ ms the temperature is about 20 MeV and
$\mathcal{R}=\mathcal{R}_q\simeq 1.5\times 10^{15}$ s$^{-1}$
fm$^{-2}$, then the strangeon nuggets with initial baryon number
$A_0\leq 10^{36}$ will almost disappear as the result of
evaporation, via emitting strangeons.
Because the initial baryon numbers of ejected strangeon nuggets are
between 1 to $10^{39}$, we can infer that if the initial temperature
is about 20 MeV, then almost all of the ejected nuggets turn into
strangeons within several milliseconds.

In the spiral arms from tidal interactions during the merger,
however, the temperature would not be so high.
Below 10 MeV, the emissions of neutrons and protons will be dominant
instead of strangeons, which would happen in the spiral arms in the
equatorial plane.
Eq.(\ref{A}) indicates that, if the time duration from $T\sim 10$
MeV to 5 MeV is about 10 ms, when $\mathcal{R}=\mathcal{R}_n\simeq
10^{10}$ s$^{-1}$ fm$^{-2}$, then the strangeon nuggets with initial
baryon number $A_0\leq 10^{24}$ will almost disappear within tens of
milliseconds as the result of evaporation, via emitting neutrons and
protons.

\section{Strangeon kilonova}
\label{sec:strangeon kilonova}

The scenario of strangeon kilonova was proposed in Paper I, where
the light curves are powered by the decay of ejected strangeon
nuggets and the spin-down of the remnant strangeon star.
To be consistent with observations of the kilonova AT 2017gfo
following GW170817 \citep{Kasliwal2017}, the lifetime of the
strangeon nuggets was assumed to be 1 day.
Here we propose a more reasonable scenario of strangeon kilonova
based on a more detailed analysis of ejected strangeon nuggets and
their evolutions.

In Section~\ref{sec:evaporation} we discuss a possible ejection
process of merging double strangeon stars.
The merger ejects strangeon nuggets directly, which would suffer
from evaporation of particles, mainly strangeons, neutrons and
protons.
The tidal disruption during the merger ejects strangeon nuggets in
the equatorial plane, which turn into neutrons and protons within
dozens of milliseconds.
The hydrodynamical squeeze from the contact interface between the
merging stars ejects strangon nuggets in a broad range of angular
directions, which turn into strangeons within several milliseconds.

\subsection{Electron fraction}
\label{subsec:electron fraction}

Strangons are unstable and will decay instantly.
Although we do not know the exact microscopic properties of a
strangeon, we could infer its decay channels by analogy with
hyperons.
Taking $\Lambda$ hyeron as an example.
Its lifetime is $\sim 10^{-10}$ s and the decay channels
are~\citep{PDG2018}
\begin{eqnarray}
\Lambda \longrightarrow p+\pi^-\ \ (63.9\%) \label{lambda to p} \\
\Lambda \longrightarrow n+\pi^0\ \ (35.8\%) \label{lambda to n}
\end{eqnarray}
The produced $\pi^-$ and $\pi^0$ are still short-lived with
lifetimes $\sim 10^{-8}$ s and $\sim 10^{-17}$ s respectively, and
will decay via~\citep{PDG2018}
\begin{eqnarray}
\pi^- &\longrightarrow & \mu^- +\nu_{\mu} \label{pi to mu} \\
\mu^- &\longrightarrow & e^- +\nu_{\mu}+\bar{\nu_e} \label{mu to
bar-nu-e}
\end{eqnarray}
and
\begin{eqnarray}
\pi^0 &\longrightarrow & 2\gamma \ (98.82\%) \\
\pi^0 &\longrightarrow & e^+ +e^- +\gamma \ (1.17\%)
\end{eqnarray}
Therefore, we infer that the main decay products of strangeons would
also be protons, neutrons, e$^-$, $\bar{\nu_e}$, $\nu_\mu$ and
photons.

It is interesting to note that, in evaporation of strangeon nuggets
and decay of strangeons, the ratio of production rate of neutrons to
that of protons are different.
In the evaporation products of strangeon nuggets, the neutrons
dominant over protons, since the emission of protons is suppressed
due to the Coulomb barrier.
However, in the decay products of strangeons, there are more protons
than neutrons, since protons are lighter than neutrons.
This difference basically initiates different levels of
neutron-richness in the ejecta that will be discussed later.

In summary, the strangeon nuggets ejected directly from the merger
would emit particles from the surface, which are dominated by
strangeons at $T>10$ MeV and neutrons at 1 MeV $<T<10$ MeV.
%
Strangeons are extremely unstable and will instantly decay into
proton-rich matter, so electron fraction $Y_{\rm e}$ of the ejecta
depends on temperature.
Taking into account both the emission rates derived in
Sec~\ref{subsec:decay widths} and the decay of strangeons, we can
get the dependence of $Y_{\rm e}$ on temperature, as shown in Fig.2.
We can see that, $Y_{\rm e}$ is higher than 0.5 at $T>10$ MeV and is
well below 0.1 at 1 MeV $<T<10$ MeV.
%
   \begin{figure}[h!!!]
  \centering
   \includegraphics[width=9.0cm, angle=0]{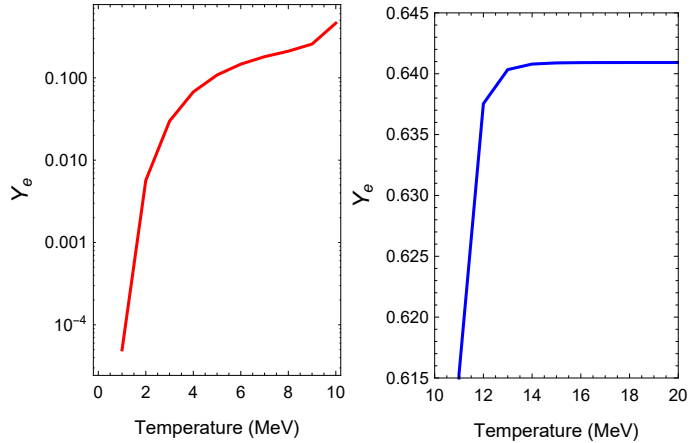}
   \begin{minipage}[]{85mm}
   \caption{The dependence of electron fraction $Y_{\rm e}$ on temperature $T$ in the ejecta,
   taking into account both the evaporation of strangeon nuggets and the decay of strangeons.}
   \end{minipage}
   \label{fig_Ye-T}
   \end{figure}

\subsection{Two-component ejecta}

The ejection processes of strangeon nuggets involves the tidal
disruption that ejects matter in the equatorial plane, and the
hydrodynamical squeezing from the contact interface between the
merging stars that expels matter in a broad range of angular
directions.
Therefore, as the temperature would be different in different
processes, the neutron-rich matter would be ejected from the
directions around the equatorial plane, and the proton-rich matter
would be ejected in a broad range of angular directions.
All of the above processes happen within tens of milliseconds.

Consequently, we may infer that the end products of the complex
interactions within tens of milliseconds from the merger of double
strangeon stars could be similar to that ejected in the merger of
double neutron stars.
In other words, after about dozens of milliseconds from the
coalescence, the ejecta of merging double strangeon stars could be
similar to that of merging double neutron stars, both of which would
power the kilonova-like transient.

The neutron-abundance of ejecta depends on the viewing angles.
Besides the emitted neutrons from strangeon nuggets that make the
equatorial plane neutron-rich, the strangeons emitted from strangeon
nuggets could also contribute to the neutron-richness.
In the high density region of the disk, the produced $\bar{\nu_e}$
in decay (\ref{mu to bar-nu-e}) would transform protons into
neutrons, via $p+\bar{\nu_e} \longrightarrow n+e^+$.
Anyway, the matter ejected from around the equatorial plane could be
neutron-rich.
Moreover, even if the remnant is a long-lived stable star, the
radiation from the star would be insufficient to increase $Y_{\rm
e}$ significantly, since most of the ejecta in the equatorial plane
can have very low $Y_{\rm e}$, indicated in Fig.2.

The components of ejecta are illustrated in Fig.3.
The tidal disruption ejects matter in the equatorial plane, where
the temperature is relatively low.
The hydrodynamical squeeze from the contact interface expels matter
in a broad range of angular directions, where the temperature is
relatively high.
Therefore, taking into account both the evaporation of strangeon
nuggets and the decay of strangeons, the matter with high opacity
would be ejected from the directions around the equatorial plane,
and the matter with low opacity would be ejected in a broad range of
angular directions.

   \begin{figure}[h!!!]
  \centering
   \includegraphics[width=9.0cm, angle=0]{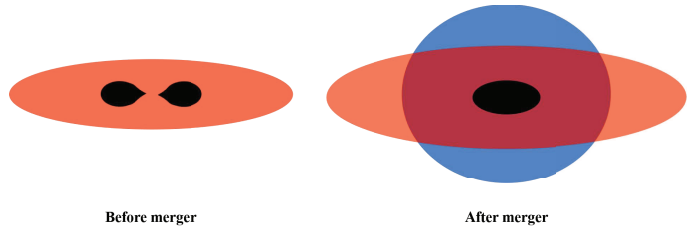}
   \begin{minipage}[]{85mm}
   \caption{Illustration of blue and red components of ejecta.
   The matter with high opacity would be ejected from the directions around the equatorial plane,
   and the matter with low opacity would be ejected in a broad range of angular directions.}
   \end{minipage}
   \label{fig_blue_red}
   \end{figure}

\subsection{Light curves}

To derive the light curve, the radiation-transfer process is the
necessary input.
As demonstrated before, the ejecta after tens of milliseconds after
merger could be similar to that ejected in the merger of double
neutron stars.
The neutron-rich matter (i.e. the red component), would be ejected
from the directions around the equatorial plane, and the proton-rich
matter (i.e. the blue component) would be ejected in a broad range
of angular directions.
The r-process nuclei can be produced in the neutron-rich
environment, leading to high opacity and heat the ejecta by
radioactive decay.
Therefore, the radiation-transfer process would be similar to that
of merging double neutron stars.
The difference is that, the amount of heavy nuclei produced in
merging strangeon stars would be much smaller than that in merging
neutron stars, since the total ejected mass of the former would be
much smaller than that of the latter.

The maximum mass of strangeon stars would be as high as
$2.3M_{\odot}$ or even higher, so the merger of double strangeon
stars triggering GW170817 would probably left a long-lived stable
strangeon star.
As indicated in~\citet{LiYu2018}, the emission of AT2017gfo
associated with GW170817 can be explained by energy injection from a
long-lived and spinning-down neutrons stars.
The spin-down power is independent of the interior structure of the
remnant, so we can take the spin-down power as the energy source of
the kilonova-like transients.
The radiation-transfer process depends on properties of the ejecta,
such as the total mass $M_{\rm ej}$, the minimum and maximum
velocities $v_{\rm min}$ and $v_{\rm max}$, the density distribution
index $\delta$ and the opacity $\kappa$.
Here we choose typical values for such parameters.
For both blue and red component, $M_{\rm ej}=10^{-4}M_{\odot}$,
$v_{\rm min}=0.1c$, $v_{\rm max}=0.3c$ ($c$ is the speed of light),
and the density distribution index $\delta=1.5$.
The opacity $\kappa=0.1$ cm$^2$ g$^{-1}$ for blue component, and
$\kappa=20$ cm$^2$ g$^{-1}$ for red component, respectively.
In order to significantly spin down the remnant, efficient secular
GW spin-down is needed.
The timescale of spin-down is $t_{\rm sd}=3\times 10^3$ s, and the
initial spin-down luminosity is $1.5 \times 10^{43}$ erg, which are
typical values for spinning down neutron stars.

Bolometric light curve of a strangeon kilonova including
two-component ejecta, fitted to the data from~\citet{Kasliwal2017}
are shown in Fig.4.
The dashed and dash-dotted lines represent the light curves of blue
and red components, respectively.
The solid line is the result of the combination of the two
components.

   \begin{figure}[h!!!]
  \centering
   \includegraphics[width=9.0cm, angle=0]{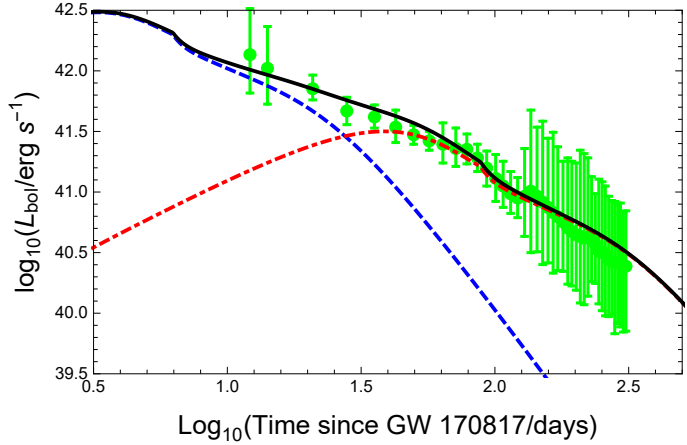}
   \begin{minipage}[]{85mm}
   \caption{Bolometric light curve of a strangeon kilonova including two-component ejecta, fitted to the data from~\citet{Kasliwal2017}.
   The dashed and dash-dotted lines represent the light curves of blue and red components, respectively.
   The solid line is the result of the combination of the two components.
   For both blue and red component, the ejected mass $M_{\rm ej}=10^{-4} M_{\odot}$,
   the minimum and maximum velocities $v_{\rm min}=0.1c$ and $v_{\rm max}=0.3c$ ($c$ is the speed of light),
   the density distribution index $\delta=1.5$.
   The timescale of spin-down is $t_{\rm sd}=3\times 10^3$ s, and the initial spin-down luminosity is $1.5 \times 10^{43}$ erg s$^{-1}$.
   The opacity $\kappa=0.1$ cm$^2$ g$^{-1}$ and 20 cm$^2$ g$^{-1}$ for blue and red components, respectively.
   The details of the ejecta model are given in~\citet{Yu2017}.}
   \end{minipage}
   \label{fig_lc_blue_red}
   \end{figure}

Therefore, although the very initial components in ejecta of merging
strangeon stars are different from that of merging neutron stars,
the ``strangeon kilonova'' could have light curves similar to that
of neutron kilonova.
Under reasonable values of parameters, the bolometric light curve
can fit the data of AT2017gfo.

\section{Conclusions and discussions}
\label{sec:conclusions}

Strangeon matter in bulk is conjectured to be more stable than
nuclear matter, and strangeon stars are conjectured to be actually
pulsar-like compact stars.
Besides the strangeon stars that born in supernova explosions and
undergo sufficient cooling, the astrophysical consequences in the
hot environment created by merging double strangeon stars are worth
exploring, especially in the new era of multi-messenger astronomy.
To develop the ``strangeon kilonova'' scenario proposed in Paper I,
we make a qualitative description about the evolution of ejecta and
light curves of strangeon kilonova.

Due to the self-bonding of strangeon stars, the merger directly
ejects strangeon nuggets instead of individual strangeons.
The tidal disruption ejects strangeon nuggets in the equatorial
plane, and the hydrodynamical squeeze from the contact interface
expels strageon nuggets in a broad range of angular directions.
In the high temperature environment of the merger, the ejected
strangeon nuggets would suffer from evaporation into strangeons,
neutrons, protons and so on.
The emission of strangeons dominates at temperature above $\sim 10$
MeV, and the emission of neutrons dominates at temperature blew
$\sim 10$ MeV.

The temperature of the matter expelled by hydrodynamical squeeze
from the contact interface could be higher than 10 MeV, so the
evaporation productions are dominated by strangeons, and almost all
of the ejected nuggets turn into strangeons within several
milliseconds.
Strangeons in free space are extremely unstable and would
immediately ($\sim 10^{-10}$ s) decay, and the decay products would
contain more protons than neutrons.
Besides, the temperature in the spiral arms from tidal interactions
would be around or below 10 MeV, but would last for relatively
longer timescale of tens of milliseconds, which still lead to
sufficient evaporation, and the evaporation productions are
dominated by neutrons.

Taking into account both the evaporation of strangeon nuggets and
the decay of strangeons, we find that, the neutron-rich matter would
be ejected from the directions around the equatorial plane, and the
proton-rich matter would be ejected in a broad range of angular
directions.
The r-process nuclei can be produced in the neutron-rich
environment, leading to high opacity and heat the ejecta by
radioactive decay.
Therefore, the radiation-transfer process would be similar to that
of merging double neutron stars.

We find similarities between the consequences of merging strangeon
stars and that of merging neutron stars, although the very initial
components in ejecta of the former are different from that of the
latter.
Light curves are then for both low and high opacity components,
under typical model of ejecta to include the radiation-transfer
process.
Under reasonable values of parameters, even if the ejected mass is
only $\sim 10^{-4} M_{\odot}$, the bolometric light curves can fit
the data of AT2017gfo, by the energy injection from a long-lived and
spinning-down strangeon star.
Although the rotational energy released by the remnant during its
spin-down will be transferred into the GRB jet~\citep{Margalit2017},
the radiation of the fast rotating remnant would be compatible with
the observed GRB if the magnetic field of the remnant is not higher
than $10^{12}$ Gauss~\citep{Yu2017}.

This paper is the first qualitative description about the evolution
of ejecta of merging strangeon stars.
Despite our lack of numerical simulations, our conclusions are
qualitatively acceptable, for the following reason.
Most of the ejected strangeon nuggets would almost disappear and
evaporate into strangeons, neutrons and protons, then strangeons are
instantly decay into protons and neutrons.
The ejection, evaporation and decay happen at very early stage of
merger and terminated at time about tens of milliseconds when
temperature dropped below $\sim 1$ MeV, so that the ejecta would end
up with neutrons and protons within tens of milliseconds.
Consequently, the early processes could not have much impact on the
later processes such as the r-process nucleosynthesis and strangeon
kilonova.
Future numerical simulations are necessary to explore the full
processes and consequences of merging double strangeon stars.

How to distinguish strangeon stars and neutron stars by the
observational consequences is crucial to test the strangeon star
model.
We find that, even if the remnant is a long-lived stable star, the
radiation from the star would be insufficient to increase $Y_{\rm
e}$ significantly, since most of the ejecta in the equatorial plane
can have $Y_{\rm e}$ well below 0.1 (lanthanide-bearing).
As found in our previous work, the merger of double strangeon stars
triggering GW170817 would probably left a long-lived stable
strangeon star.
Therefore, the merging strangeon stars scenario seems to be helpful
to include both long-lived remnant and sufficient lanthanide-bearing
ejecta.
Conversely, for merging double neutron stars, most of the ejecta
would have $Y_e\gtrsim 0.3$ (lanthanide-free) if the remnant
survives longer than about 300 ms~\citep{Kasen2015}.
More information about the post-merger remnant of GW170817 in the
future will undoubtedly provide more sever test for both neutron
star and strangeon star models.

The above statements are based on the hypothesis that, the the
emission of neutrinos of newly born strangeon stars are the same as
that of newly born neutron stars, in which case the luminosity of
$\nu_e$ is larger than that of $\bar{\nu_e}$.
The emission of neutrinos of newly born strangeon stars is still
unknown, so the consequences of neutrino radiation from the hot
strangeon stars on the ejacta and torus remain to be answered.
It is interesting to see that, neutrinos could be a probe to
distinguish strangeon stars and neutron stars, if the decay of
strangeons is similar to that of hyperons.
As indicated in~\ref{subsec:electron fraction}, the decay of
strangeons would produce a large amount of $\nu_\mu$, which would
not be produced as much in neutron star mergers.
This may be tested by neutrino detections, e.g. the Super-Kamioka
Neutrino Detection Experiment.

The critical baryon number $A_{\rm c}$ of stable strangeon nuggets,
smaller than which the strangeon matter will decay to $ud$ matter,
should be determined by both the weak and strong interactions.
The value $10^9$ for $A_{\rm c}$ adopted in
Section~\ref{sec:ejection}, by setting the critical size to be the
Compton wavelength of electrons, is actually determined by the weak
interaction only.
If the strong interaction dominates, $A_{\rm c}$ could be much
smaller, e.g., the calculations under a liquid drop model show that
$A_{\rm c}$ could be as low as $\sim 10^3$~\citep{Wangzhen2018}.
Consequently, the actual value of $A_{\rm c}$ might be in the range
from $10^3$ to $10^9$.
Certainly, the exact value of $A_{\rm c}$ would not affect the
physical picture concerned in this paper.

It is worth noting that, the consequence of survived nuggets would
also be interesting.
In calculating the evaporation rate of strangeon nuggets, we neglect
the surface tension, since larger nuggets would emit more particles
and we only care about the emitted particles that affect the
subsequent transient, then we find that a small amount of large size
nuggets, with initial baryon number $A_0>10^{24}$ produced by tidal
forces and $A_0>10^{36}$ produced by hydrodynamical squeeze, can
survive evaporation.
However, when the radius of a nugget decrease to $\sim 10^5$ fm
(with baryon number $\sim 10^{15}$), the surface tension would
become significant, which would lower the emission rate and make it
easier to survive.
Moreover, although most of the baryons are lost during evaporation,
the absorption of energy and decrease of temperature due to
evaporation may prevent further evaporation, then the strangeon
nuggets with smaller $A_0$ may left to be microscopic strangeon
nuggets with $A\gtrsim A_{\rm c}$.
The survived strangeon nuggets would perform like the ultra high
energy cosmic rays, and their density in galaxies and impact on the
evolution of stars are worth exploring in the future.

\normalem
\begin{acknowledgements}
We would like to thank Dr. Shuang Du (PKU) for useful suggestions.
This work is supported by National SKA Program of China No.
2020SKA0120300, the National Key R\&D Program of China (No.
2017YFA0402602), the National Natural Science Foundation of China
(Nos. U1831104, 11673002, U1531243, 11705163, 11822302, 11803007),
the Strategic Priority Research Program of CAS (No. XDB23010200),
and Ningbo Natural Science Foundation (Grant No.~2019A610066). The
support provided by China Scholarship Council during a visit of
C.-J. X to JAEA is acknowledged.
\end{acknowledgements}

\bibliographystyle{raa}
\bibliography{reference}

\label{lastpage}

\end{document}